\documentclass[final,5p,times,twocolumn]{elsarticle}
\usepackage{graphicx}
\usepackage{hyperref}
\usepackage{amssymb}
\usepackage{amsthm}
\usepackage{forest}

\newcounter{bla}

\journal{Computer Physics Communications}

\begin{document}
\begin{frontmatter}
\title{\texttt{Imeall}: A Computational Framework for the Calculation of the Atomistic Properties of Grain Boundaries}

\author[a]{H. Lambert \corref{author}}
\author[a]{Adam Fekete}
\author[b]{J. R. Kermode}
\author[a,c]{A. De Vita}

\cortext[author]{Henry Lambert \\\textit{E-mail address:} henry.lambert@kcl.ac.uk}
\address[a]{Department of Physics, King's College London, Strand, London, WC2R 2LS, United Kingdom} 

\address[b]{Warwick Centre for Predictive Modelling, School of Engineering, University of Warwick, Coventry CV4 7AL, United Kingdom}
\address[c]{Department of Engineering and Architecture, University of Trieste, I-34127 Trieste, Italy}
\begin{abstract}
We describe the \texttt{Imeall} package for the calculation and indexing of atomistic properties of grain boundaries in materials. 
The package provides a structured database for the storage of atomistic structures and their associated properties, equipped with
a programmable application interface to interatomic potential calculators.
The database adopts a general indexing system that allows storing
arbitrary grain boundary structures for any crystalline material.
The usefulness of the \texttt{Imeall} package is demonstrated by computing, 
storing, and analysing relaxed grain boundary structures for a 
dense range of low index orientation axis symmetric tilt and twist boundaries in $\alpha$-iron for
various interatomic potentials. 
The package's capabilities are further demonstrated by carrying out automated structure generation, dislocation analysis, interstitial site detection, 
and impurity segregation energies across the grain boundary range.   
All computed atomistic properties are exposed via a web framework, providing open access to the grain boundary repository and the analytic tools suite.
\end{abstract}

\end{frontmatter}

{\bf PROGRAM SUMMARY}
\begin{small}
\noindent
{\em Program Title: \texttt{Imeall}}                             \\
{\em Licensing provisions: Apache-2.0} \\
{\em Programming languages: python, fortran, javascript}                   \\
{\em Nature of problem:}\\
 Determining the minimum energy structure for a specific grain boundary 
and interatomic force field involves extensive searches in configuration space. 
 Any duplication of this effort should be avoided, and providing a unique database
to gather the resulting structures is needed for this.    
  Accurately cataloguing the chemical and electronic environments
associated with the interface atoms in the grain boundary furthermore requires a
usable interface between a database of grain boundary structures 
and an expandible set of interatomic potential calculators.\\
{\em Solution method:}\\ 
 We introduce a standard indexing convention that allows the integration of grain boundary structures for arbitrary materials, 
 generated by different users/research groups, into a normalised database that can be easily queried and used as a starting 
 point for further research projects. 
The \texttt{Imeall} package accomplishes this by specifying a standard naming convention for the grain boundary database
and by providing the software routines necessary to populate and query such a database, as well as an interface to interatomic calculators and analysis tools.\\
\\

\end{small}

\section{Introduction}
\label{sec:intro}
A grain boundary is formed wherever an extended planar region can be identified that separates two
crystalline regions of a homophase material differing in the relative orientation of the adjoining crystal lattices.
Grain boundaries play a significant role in determining the mechanical properties of materials. 
Having access to a standardised database of grain boundary structures
is an important prerequisite for any atomic-scale (``chemo-mechanical") analysis, e.g., 
to investigate how impurities trapped at grain boundaries influence the mechanical strength of the material.  
Evaluating several properties of significant engineering interest requires access to the
atomistic structures of grain boundaries. 
These include, but are by no means limited to, the diffusivity of impurities at 
grain boundary interfaces Ref.~\cite{vattre16}, the segregation energies of 
impurities to interfaces Ref.~\cite{zhou17}, and the interaction 
and slip transmission of dislocations across boundaries Refs.~\cite{bachurin10, spearot14}.
While a number of systematic investigations of grain boundary structures have been performed, 
see Refs.\cite{bristowe75, vitek80, tschopp07} and references therein, a single repository equipped 
with the appropriate tools to generate, archive and analyse the boundaries is still missing. 
The \texttt{Imeall} package and the suite of routines described in this work 
make a step in this direction by providing a framework for constructing 
and cataloguing grain boundary structures, determining their minimum energy, 
and calculating an array of quantities of interest.

The paper is organised as follows. 
Sec.~\ref{sec:structgeneration} describes 
the procedure and routines for systematically generating tilt and twist grain boundaries.
Sec.~\ref{sec:gbdb} introduces a unique naming convention which is
reflected in the structure of the \texttt{Imeall} repository. 
This scheme allows the database to incorporate grain boundary
structures for any reference crystalline material, computed using any interatomic potential,
in a consistent and physically intuitive fashion.
The information contained in the database is mirrored by a normalised SQL database to allow rapid queries
via the command line or a web interface. 
Sec.~\ref{sec:packagelayout} describes the overall layout of the computational package, highlighting the 
key routines needed to perform the atomistic calculations and analysis, and those used for storing this information in the database and exposing it via a web framework.
The following sections describe some applications of the package,
to illustrate the comparative information that can be exposed.  
In particular, Sec.~\ref{sec:gbrelax} uses the package to investigate how grain boundary
energetics and structural topologies vary upon using different
interatomic potential parameterisations. 
Finally, Sec.~\ref{sec:interstitials} describes how the package handles calculations 
addressing point-defect energetics, specifically interstitial hydrogen located at bcc 
Fe grain boundaries.  

\section{Generating Grain Boundary Structures}
\label{sec:structgeneration}
The full macroscopic specification of a grain boundary can be accomplished using
five degrees of freedom \cite{sutton95}.
To generate a boundary, two identical crystals, which we 
may here differentiate by referring to them as the 
`red crystal' and the `blue crystal' (see Fig. 1) 
are initialised in the same orientation.  
The red crystal is then rotated by an angle $\theta$ (a single degree
of freedom) around a rotation axis specified by the vector $\hat{N}$
(associated with two more degrees of freedom).  
After this rotation, a boundary plane is chosen, specified by its
normal vector $\hat{bp}$ expressed in the unrotated coordinate system. 
This final vector exhausts the remaining two degrees of freedom and concludes
the macroscopic specification of the grain boundary. 
All red crystal atoms located below the boundary plane and all blue
crystal atoms located above the boundary plane are at this
point removed.  
We will refer to the choice of orientation axis, misorientation angle, and 
boundary plane as a complete \textit{canonical} macroscopic specification of the grain boundary, 
i.e., before atomistic relaxations are considered.  
It is useful for what follows to define two grain boundary main geometries. 
If the rotation axis is parallel to the boundary plane normal  ($\hat{N}$ parallel to $\hat{bp}$ ),
the grain boundary is referred to as a twist boundary.
If the rotation axis is orthogonal to the boundary plane normal ( $\hat{N}$ orthogonal to $\hat{bp}$ ), 
the grain boundary is referred to as a tilt boundary.
In the case where the Miller indices of the boundary plane are 
the same in the coordinate systems of both grains, the 
grain boundary is referred to as symmetric.
In the \texttt{Imeall} package generating an exhaustive array of tilt 
symmetric and twist grain boundary structures is accomplished by the methods defined in
\texttt{imeall.slabmaker.slabmaker} and \texttt{imeall.slabmaker.gengb\_from\_quat}.
These routines make use of quaternion algebra to systematically generate a full
array of symmetric tilt and twist boundaries (see Refs.~\cite{handscomb58, zeiner05}
for other applications of quaternion algebra to the study of grain
boundaries; an overview of quaternion algebra is given in Ref.~\cite{goldman10}). 
Quaternions are frequently used in engineering and graphic design contexts 
to handle vector rotations. 
For the present purposes, quaternions can be thought of as four dimensional
objects with one scalar component and a three-dimensional vector
component. 
The rotation of a vector  $\hat{v}$  by an angle $\theta$ around a unit vector $\hat{N}$ is accomplished 
by conjugation of  $\hat{v}$ by the quaternion $q$, i.e., by multiplying the vector on the left and the right 
as $q \hat{v} q^{-1}$, where $q^{-1}$ denotes the inverse of $q$.
In the context of grain boundaries the four components of the quaternion, 
$q=(\theta, \hat{N})$ are converted to
$q=(\cos(\theta/2), \sin(\theta/2)\hat{N})$ so that the necessary rotations 
may be obtained. 
These components are directly obtained
from the physical parameters defining the grain boundary i.e. the misorientation angle
and the orientation axis.
Interestingly, if the quaternion $q$ can be reduced to a {\em primitive form}, i.e. all its entries 
are integers apart from a scaling factor, the rotation is guaranteed 
to produce a coincident site sublattice (cf. Refs.~\cite{grimmer74,
grimmer84}) that is, a periodic sublattice of space points where red and blue atoms 
coincide.  
In this case, given $q$ , closed expressions exist that readily provide
a set of basis vectors for the coincident site sublattice Ref.~\cite{zeiner05}.

\begin{figure}[htb]
\begin{center}
\includegraphics[width=0.8\columnwidth]{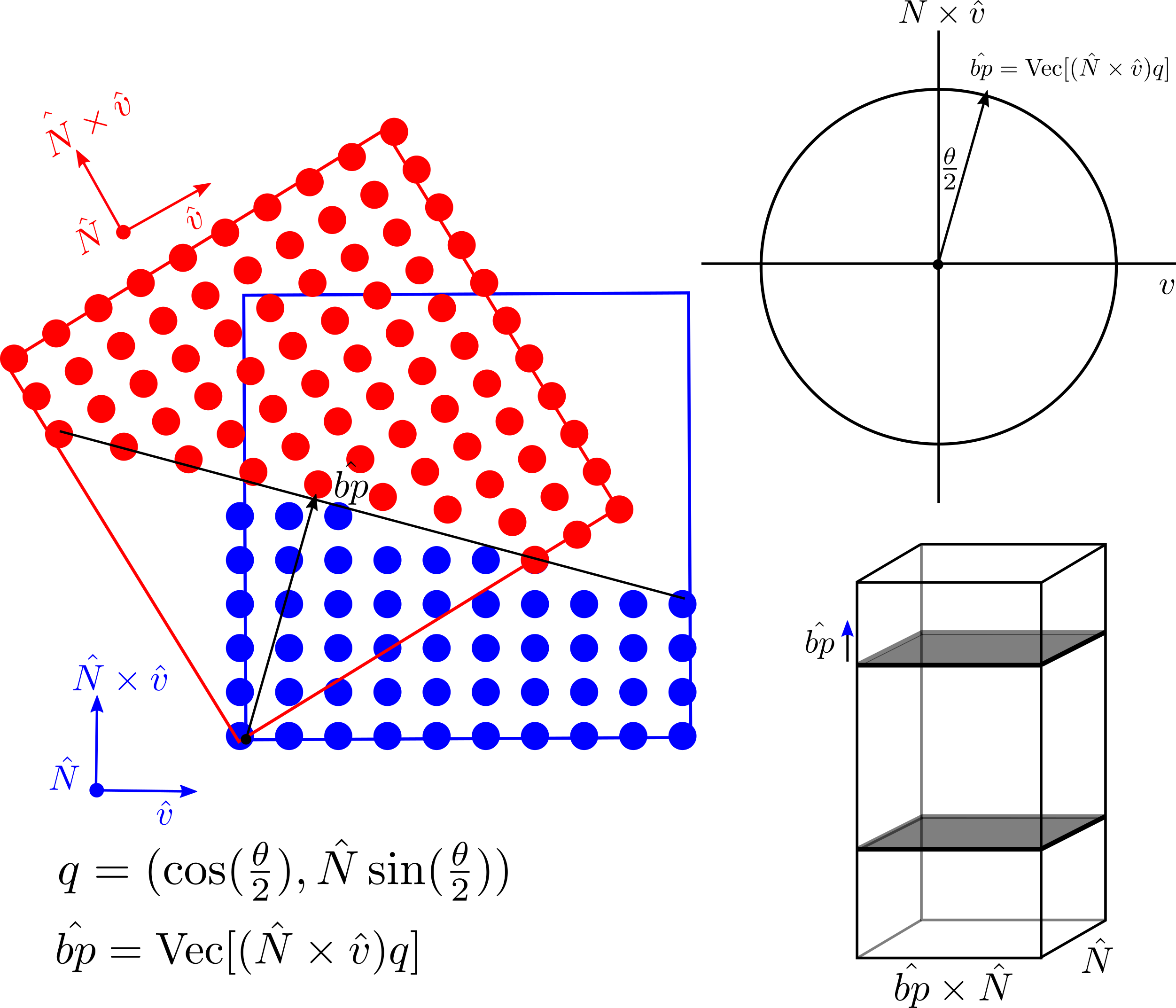}
\caption{The general coordinate system for determining tilt and twist boundary planes and 
orientations using quaternion algebra. 
Hats indicate regular three dimensional vectors, 
considered to be quaternions with a 0 component for the scalar part. 
The quantity q, without a hat, represents a quaternion with a vector
part determined by the orientation axis $\hat{N}$ and a
scalar part determined by the misorientation angle $\theta$. 
The effect of quaternion multiplication on the vectors normal to the 
orientation axis plane are represented schematically.
Upper right inset: the total relative rotation $\theta$ of the 
two misoriented grains is depicted by the red and blue
crystals with their coordinate systems rotated. 
Lower right inset: the bicrystal generated by reflecting
the grain boundary to double the number of grain boundaries in a unit cell. 
This guarantees periodicity perpendicular to the grain boundary plane
allowing for the use of periodic boundary conditions in the atomistic relaxation.\label{fig:gbgen}}
\end{center}
\end{figure}

Besides making it easy to perform rotations with no need for using 
rotation matrices, there is an additional favourable aspect of the quaternion 
scheme for generating symmetric tilt grain boundaries.
This is illustrated in Fig.~\ref{fig:gbgen}, which presents a schematic representation for the determination of 
the boundary planes for a tilt symmetric grain boundary using quaternion algebra. 
The two crystals, red and blue, are initially superimposed. 
In Fig.~\ref{fig:gbgen} the unit vector denoting the axis where the misorientation
angle is measured from, according to a right hand rule, is denoted $\hat{N}\times\hat{v}$.
Here we have assumed that $\hat{N}$ is normal to a lattice plane (the
coordinates of $\hat{N}$ will be thus three integers apart from a scaling factor) 
and $\hat{v}$ is a lattice vector of the plane, orthogonal to $\hat{N}$. 
The right quaternion (``half rotation'') product $(N\times v)q$ can be used in this geometry 
to define the boundary plane normal which in the original reference frame 
corresponds to the symmetric tilt grain boundary plane, i.e., the grain boundary 
associated with the misorientation angle $\theta$  and axis $\hat{N}$ encoded 
as components of $q$. 
All red lattice vectors $\hat{\tilde{v}}$  can be obtained at this point
by a left and right quaternion product (``full rotation"), as $q \hat{\tilde{v}} q^{-1}$. 
Once all atomic positions of the grain boundary interface have been generated, 
the unit cell is doubled by reflecting it through the boundary plane, 
to produce a periodic structure (cf. Fig.~1). In the \texttt{imeall} package
the \texttt{gen\_sym\_tilt} method generates an array of angle boundary plane
pairs, and the \texttt{build\_tilt\_sym\_gb} generates the bicrystal unit cells
according to these specifications. Equivalent routines exist for twist boundary
structures.
\subsection{Microscopic Search Parameters}
\label{subsec:msp}
While the macroscopic degrees of freedom of the grain boundary are completely 
specified by the geometric considerations, determining its structurally relaxed 
atomistic structure requires a systematic search through a large space of initial 
configurations. 
This search requires initialising the misoriented grains in a reference frame 
defined by a two dimensional grid of unique positions, $\hat{x}, \hat{y}$, the rigid body
translations,
representing the in-plane translation vectors, and a lattice expansion vector $\hat{e}$, 
normal to the boundary plane, which allows for atomistic relaxation of the initial grain boundary structure. 
For further details regarding the requirements of performing this search see Ref.~\cite{rittner96} and
Refs.~\cite{tschopp07, spearot07, tschopp07III, tschopp12}. 
The mechanics of carrying out the grain boundary structural relaxation search and the routines required for
performing the atomistic relaxations are detailed in Sec.~\ref{sec:gbdb} and Sec.~\ref{sec:gbrelax}. 
The methods needed to generate the atomistic grain boundary bicrystal and the graphical representation of
coincident site lattices, are handled in the \texttt{imeall.slabmaker.slabmaker} module. 
All computations above defining the atomistic cells are performed using the \verb!quippy! library [1]. 
The resulting structures are stored as extended \texttt{.xyz} files in the grain boundary database, 
which is described in the following section. 
\section{Grain Boundary Database}
\label{sec:gbdb}
\subsection{Grain Boundary Hierarchy}
In the \texttt{Imeall} package each grain boundary is assigned an identifying label, 
an ``id" tag built up in analogy with the physical specification of the grain boundary. 
Due to the  large number of additional conventions that become necessary when discussing non-cubic systems
we have elected not to use the ``mean boundary plane" formalism for specifying the grain orientation\cite{sutton95}. 
Rather, we choose an alternative specification that arises naturally 
from the procedure described in the previous sections, 
which involves initializing two superimposed crystals, 
applying rotations via a right hand rule, and following an intuitive
geometric procedure for initializing the configuration.
For this, the orientation axis  $\hat{N}$ , the angle  $\theta$ (in degrees), and the boundary plane normal 
vector $\hat{bp}$ are serialised as absolute value integers and concatenated into a string. 
For instance, a symmetric tilt boundary specified by the orientation axis [110], with a 
misorientation angle of 13.44 degrees and separated by the $(\bar{1}\ 1\ 12)$ 
boundary plane is identified by the grain boundary id \texttt{11013441112}. 
The grain boundary at this description level is referred to as the `canonical grain'.
The database is constructed by adopting a nested hierarchy starting from the 
common root position, which is followed by the material directory, the orientation axis directory, 
and finally the canonical grain id. 
As an example, assuming that the material is $\alpha$-Fe and the database root 
is stored at \texttt{`/'}, the canonical grain mentioned above
would be accessed at \texttt{`/alphaFe/110/11013441112'}. 
\footnote{We note that this convention allows for spurious floating point degeneracy in the
naming of directories: e.g., if two grains were initialised with the same orientation axis and 
boundary plane for two different misorientation angles, say 10.34 and 103.4 degrees, 
the convention would serialise to the same canonical grain name. 
This apparent redundancy is deemed acceptable to compress 
string length of the directory names and enhance the human readability of the directory tree generated. 
Any conflict is in all cases resolved and eliminated in the 
serialised database and \texttt{gb.json} file associated with every
grain directory where the fully signed vectors and angles, 
including decimal notation, are recorded without any compression.}
To complete the canonical grain descriptor level (at which there is 
still no assumptions about the interatomic potential used for detailed atomistic characterisation), 
any data determined purely by geometric considerations, e.g. 
the orientation axis or the grain boundary $\Sigma$ number (i.e. the inverse
ratio of the number of coincident sites to lattice sites), is included in a file called \texttt{gb.json}. 
The unrelaxed canonical grain is at the same time stored as an extended \texttt{.xyz} file, 
along with an image of the grain, and a schematic vector graphics image of 
two planes of the coincident site lattice at the interface. 
Below the canonical grain descriptor level, the hierarchy proceeds in a general way intended to 
capture the full range of accessible microscopic structures. 
For this, the output of a series of subsequent microscopic 
initialisations-relaxations procedures, set up according to the requirements 
described in Sec.~\ref{subsec:msp} are stored in a unique \texttt{Potential} directory.
For example, a researcher may have studied a particular grain boundary using an embedded atom 
potential, and a tight binding model.
An appropriate layout for the output of this research, all naturally 
located below the canonical grain branching level in our database tree,  
would then be \verb!.../11013441112/EAM1/! for the first 
embedded atom potential, and \verb!.../11013441112/TB! for the 
structures computed using the tight binding model. 
Continuing down the hierarchy, the different microscopic structural 
initialisations-relaxations can be constructed and labeled 
systematically within a subdirectory structure branching underneath 
each potential directory: see Table~\ref{tab:namingconv} for a 
description of the labeling convention and an example grain specification. 
The appropriate number of microscopic initialisations-relaxations to be stored will depend on the application.
The typical pattern will involve a large screening set obtained using a computationally
inexpensive potential, with selected microscopic geometries ready to be pushed forward to 
a higher accuracy calculation scheme (e.g., density functional theory-based) 
in an appropriately labeled sister potential directory. 
The usefulness of this variable precision procedue is demonstrated by its ability to handle 
issues previously identified in Refs.~\cite{campbell93, paxton96} where classical potentials
identify non-physical minimum energy grain boundary structures.
For instance, the minimum energy structure identified by embedded atom potentials has a 
tendency to favour structures preserving the coordination number at the interface.
This propensity for over-bonding introduces the risk of creating structures with an unrealistic angular distortion. 
Quantum mechanical models can, however, restore the physically correct angular 
contributions to the total energy, and thus identify the 'true' structure, to be confirmed by experiment. 
Constructing the database in the described manner facilitates the direct 
tabulation and comparison of quantities for a definite 
structure with an exact geometric specification, i.e. a grain boundary, 
as predicted by different interatomic potential models of arbitrary accuracy. 
\begin{table}[htb]
\begin{tabular}{l l l}
\hline
\hline
Specification    & Example          & String \\
\hline
Orientation Axis & [1 1 0]          & 110 \\
Angle            &  13.44           & 1344   \\
Boundary Plane   & $(\bar{1}\ 1\ 12)$ & 1112  \\  
Supercell        & $6\times2$  & \_v6bxv2 \\
Rigid Body Translation  & [0.4, 0.2]     & \_tv0.4bxv0.2  \\
Atom Deletion criterion & 1.4\AA        & \_d1.4z \\
\hline
\hline
\end{tabular}
\caption{The directory naming convention in \texttt{Imeall} with an example grain. 
The example column has been chosen to uniquely 
specify the minimum energy structure for the [1 1 0] 13.44 $(\bar{1} 1 12)$ 
canonical grain boundary. 
The chosen interatomic potential is of the embedded atom type and is 
specified in Ref.~\cite{ramasubramaniam09}. 
The full path for this grain boundary would be
\texttt{11013441112\_v6bxv2\_tv0.4bxv0.2\_d1.4z} denoting a $6\times2$ supercell with in plane
rigid body translations of 0.4, 0.2 (referring to the fraction translation in the basis of the
lattice vectors of the primitive unit cell for an orthorhombic grain boundary), 
and an atom deletion criterion for nearest neighbours below 1.4 \AA. \label{tab:namingconv}}
\end{table}
The full microscopic initialisation is determined by the supercell size of 
the subgrain, the rigid body translations, and the atom deletion criterion distance. 
The combination of a given potential and fully specified subgrain structure is referred to
as a \texttt{SubGrainBoundary}.
\begin{figure}[htb]
\begin{center}
\begin{scriptsize}
\begin{forest}
  for tree={
    font=\ttfamily,
    grow'=0,
    child anchor=west,
    parent anchor=south,
    anchor=west,
    calign=first,
    edge path={
      \noexpand\path [draw, \forestoption{edge}]
      (!u.south west) +(3.5pt,0) |- node[fill,inner sep=0.25pt] {} (.child anchor)\forestoption{edge label};
    },
    before typesetting nodes={
      if n=1
        {insert before={[,phantom]}}
        {}
    },
    fit=band,
    before computing xy={l=12pt},
  }
[/
  [alphaFe/
    [000/]
    [001/
      [00110391110/
       [gb.json]
       [EAM\_Ram/
         [00110391110\_v6bxv2\_tv0.0bxv0.0/
           [00110391110\_v6bxv2\_tv0.0bxv0.0\_d1.4z/
            [subgb.json]
            [00110391110\_v6bxv2\_tv0.0bxv0.0\_d1.4z.xyz]
           ]
         ]
         [00110391110\_v6bxv2\_tv0.4bxv0.4/]
        ]
        [DFT/]
      ]
      [00111421100/]
    ]
    [110/]
   [111/]
  ]
]
\end{forest}
\end{scriptsize}
\caption{Snapshot of an \texttt{Imeall} directory tree. 
In this case boundaries for $\alpha$-Fe with the [001], [110], [111] orientation axis exist.  
Underneath the [001] orientation axis we show the layout for for the [001] 10.39 [1 11 0]
grain boundary. 
Two interatomic potentials have been used: the EAM parameterised in Ref.~\cite{ramasubramaniam09} 
and a density functional theory calculation.
Below the EAM directory a specific \texttt{SubGrainBoundary}, according to the naming convention 
in Table~\ref{tab:namingconv}, is represented.\label{fig:dirlayout}}
\end{center}
\end{figure}
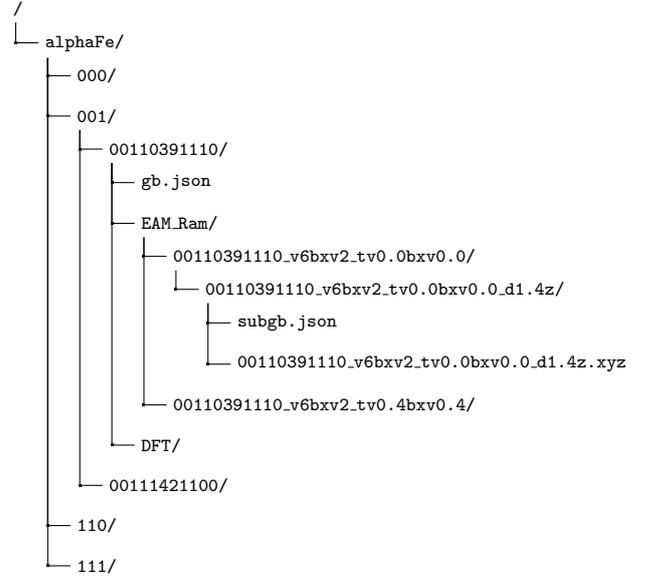
The predicted atomistic structures of a grain boundary will inevitably vary, 
depending on the interatomic potential model used 
in the calculation. 
Organising the database of microscopic structures according to potential models 
implies that the database can seamlessly 
accomodate newly predicted structures for any grain boundary, contributed from 
researchers using different interatomic potentials. 
The database also makes provisions for some extra flexibility not needed for standard grain boundaries. 
In particular, the `orientation axis' $[000]$ is reserved for isolated dislocations and fracture geometries. 
The equivalent of the canonical grain in this case specifies the type of dislocation, e.g. edge (e), screw (s),
and mixed dislocation (d), or the fracture geometry (f). 
In the case of an isolated dislocation, the Burgers vector and the dislocation line in serialised form follow 
the alphabetic specification. 
For example, an isolated edge dislocation line oriented along the [100] line with a Burgers
vector pointing along [010] would be specified as \texttt{e100010}. 
As an example of fracture geometry, \texttt{f111112}, specifies a fracture with the cleavage plane (111) 
and crack front oriented along the [1 1 -2] direction.  
Again, a \texttt{gb.json} file specifies the essential parameters of the canonical dislocation 
with subgrain database entries being resolved according to the interatomic potential used.

\subsection{Closure Tree}
\label{sec:closure}
While a hierarchical database layout is a natural choice for
organising our microscopic structure database, executing 
repeated recursive searches of the directory tree to extract 
information can become computatonally expensive. 
For this reason, the grain boundary directory hierarchy is also
`flattened' within \texttt{Imeall}, using a closure tree, that stores all
the intermediate paths to grain boundary structures, which can be rapidly queried.
This mirrored database tracks the physical database layout 
and ensures the grain boundary database remains properly normalised: i.e.
there is no duplication of canonical grain structures, or SubGrainBoundary 
objects for a given potential and there is minimal redundancy in the storage
of grain boundary properties.
Integrity keys in the database ensure a unique entry for a given
potential and a given grain boundary id. This serialised database
can also be queried rapidly without having to resort to a recursive 
search strategy. 
The \texttt{imeall.gb\_models} and \texttt{imeall.models} modules define the 
database schema for the SQLite database and provides a number of objects and
methods for rapidly querying and retrieving data. 
It is through this serialised database that the web framework for the \texttt{Imeall}
package retrieves data for the connected user. 
Using this bimodal approach to storage allows the database to coexist in the form
of an intuitive physical hierarchical layout where a researcher can manually extend 
and work within a directory tree, and a complimentary serialised 
database which can be queried in a structured and time-efficient 
manner.

\section{Package Layout}
\label{sec:packagelayout}
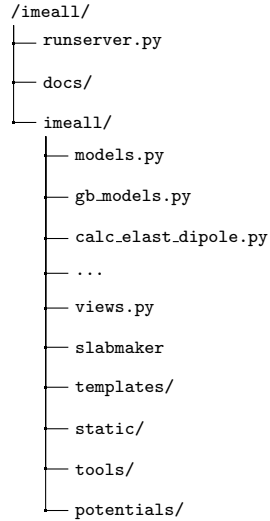
\begin{figure}[htb]
\begin{center}
\begin{scriptsize}
\begin{forest}
  for tree={
    font=\ttfamily,
    grow'=0,
    child anchor=west,
    parent anchor=south,
    anchor=west,
    calign=first,
    edge path={
      \noexpand\path [draw, \forestoption{edge}]
      (!u.south west) +(3.5pt,0) |- node[fill,inner sep=0.25pt] {} (.child anchor)\forestoption{edge label};
    },
    before typesetting nodes={
      if n=1
        {insert before={[,phantom]}}
        {}
    },
    fit=band,
    before computing xy={l=12pt},
  }
[/imeall/
  [runserver.py]
  [docs/]
  [imeall/
    [models.py]
    [gb\_models.py]
    [calc\_elast\_dipole.py]
    [...]
    [views.py]
    [slabmaker]
    [templates/]
    [static/]
    [tools/]
    [potentials/]
  ]
]
\end{forest}
\end{scriptsize}
\caption{The core directory structure and routines of the \texttt{Imeall} package.
The directories \texttt{static} and \texttt{templates}, contain the javascript and 
html templates for the web interface, the \texttt{potentials} directory contains the parameterisation
files for the interatomic potentials present in the database. 
The \texttt{models.py} and \texttt{gb\_models.py} routines define the methods and schema for traversing the
grain boundary directory tree and synchronising the directory database and the SQL database.
These routines provide the skeleton of the database. 
Routines prefixed \texttt{calc\_...} contain logic for various types of 
analysis, e.g. calculating elastic dipole tensors, performing atomistic relaxations, 
probing interstitial energetics, etc. 
The \texttt{slabmaker} module contains routines for initializing grain boundary structures.
\label{fig:packagelayout}}
\end{center}
\end{figure}

  The \texttt{Imeall} package follows the standard template for a python package.
Fig.~\ref{fig:packagelayout} demonstrates the directory structure of the package
and the key routines contained as $.py$ files.
The \texttt{Imeall} package can operate in a distributed or a local mode.
In the distributed case the host server 
can be queried and precomputed structures can be checked out or inspected via the web interface. 
Alternatively the entire \texttt{Imeall} package can be downloaded and installed by any user on a local machine, 
and, by running the \texttt{runserver.py} script a local instance of the server can be used that connects
to whatever portion of the grain boundary database the user chooses to store locally.
\section{Energetics of Relaxed Grain Boundary Structures}
\label{sec:gbrelax}
\begin{figure}[htb]
\begin{center}
\includegraphics[width=\columnwidth]{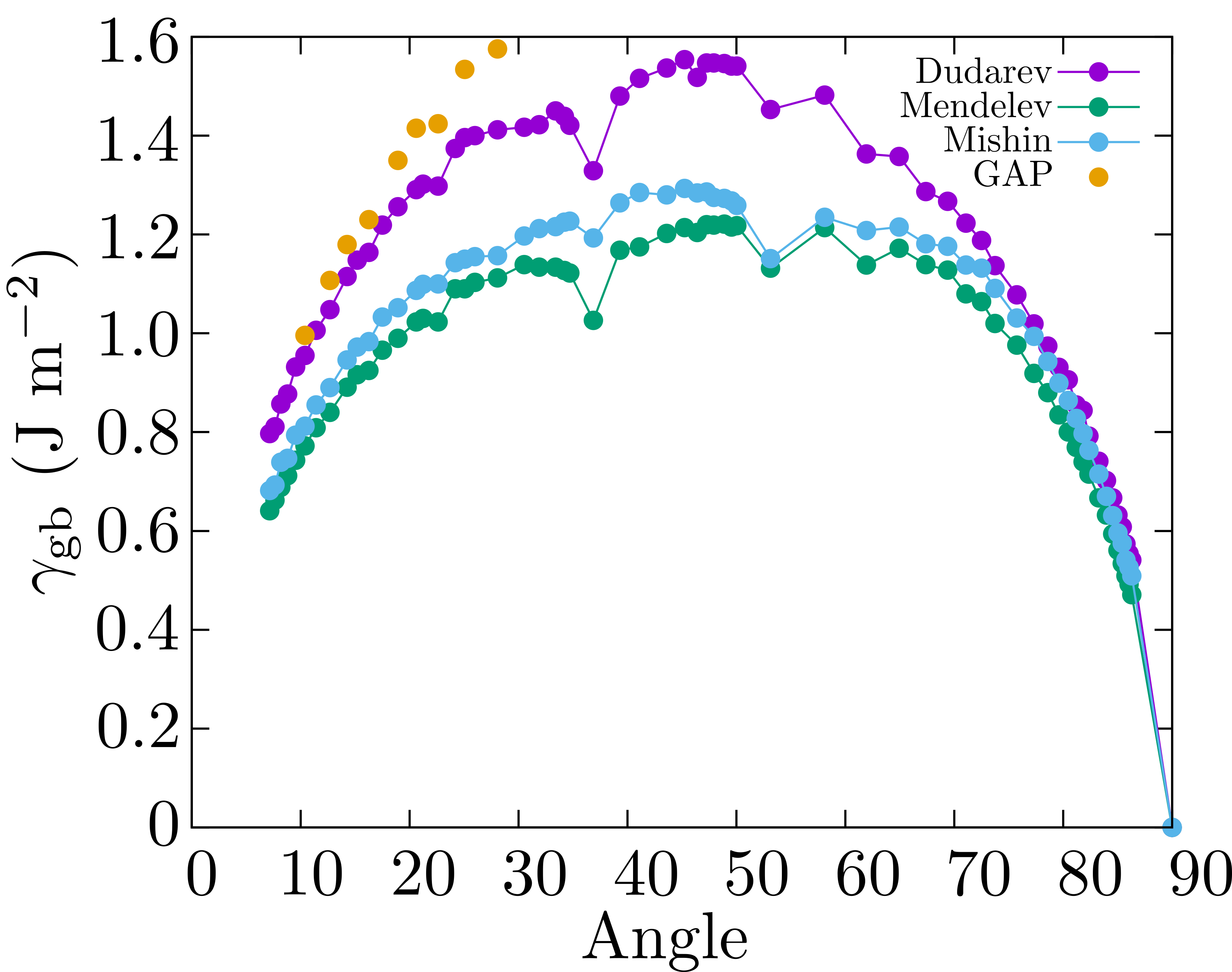}
\caption{Variation of grain boundary energies with misorientation angle for the [001]
orientation axis for a variety of embedded atom potentials and a GAP interatomic potential model.
The respective references are: Mendelev Ref.~\cite{mendelev03}, Dudarev Ref.~\cite{dudarev05}, 
Mishin Ref.~\cite{chamati06}, and using the Gaussian Approximation Potential Refs.~\cite{bartok10, bartok13}. 
The database contains the energetics for each potential along with the configuration space searched
to obtain the energetic minimum structure. The GAP model is only applied to the low angle boundaries
where there are distinct dislocations networks.
\label{fig:gb_energies}}
\end{center}
\end{figure}

\begin{figure}[htb]
\begin{center}
\includegraphics[width=\columnwidth]{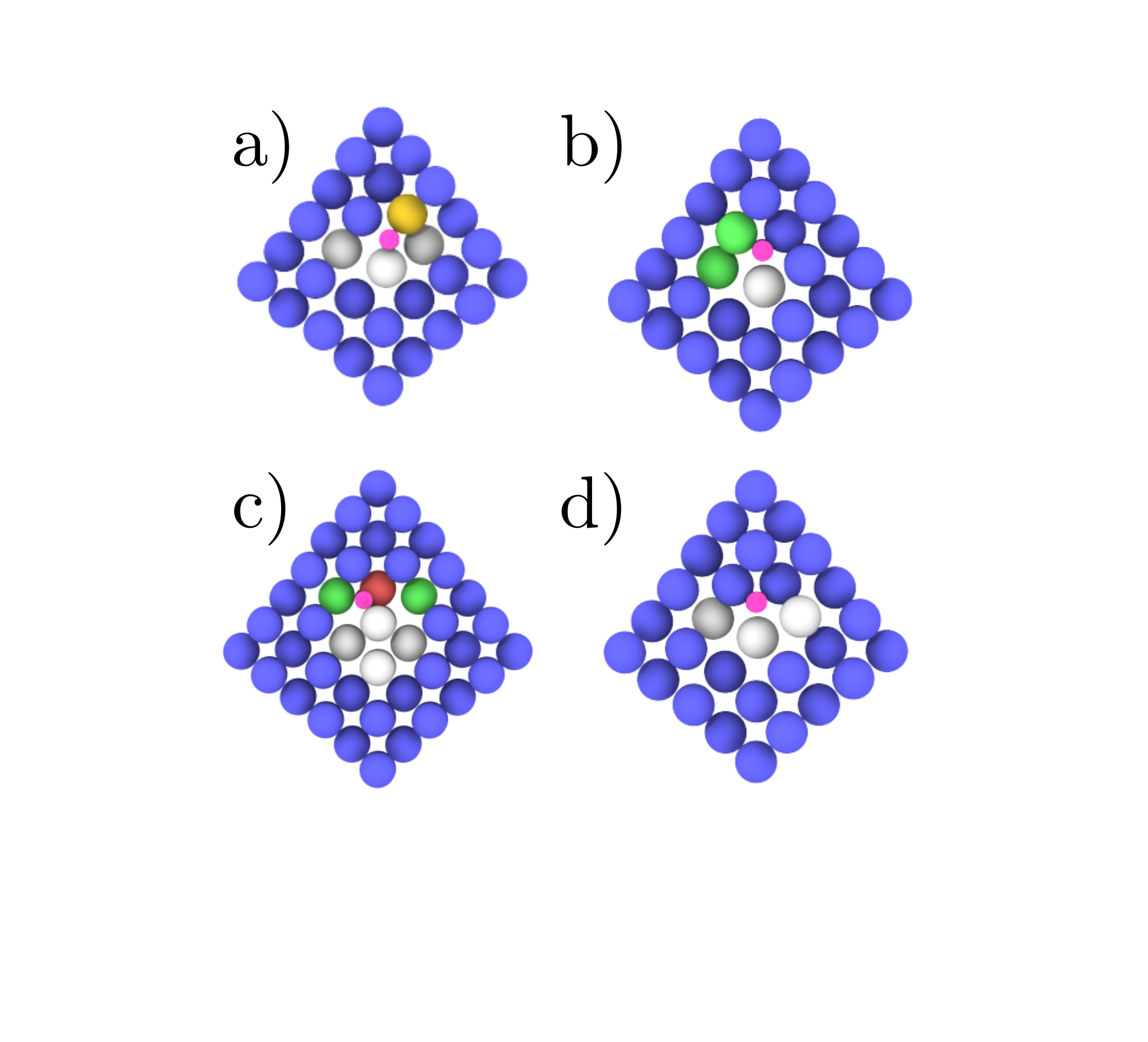}
\caption{Minimum energy dislocation cores for a grain boundary formed by a misorientation of 8.17 degrees
around the [0 0 1] axis, with a (1 14 0) boundary plane (internal imeall id \texttt{00181701140}). 
Structures have been computed using a variety of potentials. 
The colouring reflects the bond angle analysis of Ref.~\cite{ackland06} to determine
the local atomic environment. 
Namely, blue corresponds to body centered cubic, red to hexagonal close packed, green to face centered cubic, and
gold to icosahedral coordination. 
Each potential predicts the same spacing between dislocation cores, but there 
is a significant variety in the  local atomic environment  predicted by 
these potentials at each dislocation core. 
This is accompanied by a corresponding variety of the geometry of interstitial 
trap sites, which is relevant for point defect diffusivity. 
Structures a, b, c have been generated using potentials from 
Refs.~\cite{dudarev05, ramasubramaniam09, chamati06} respectively, structure d
was determined using a Gaussian approximation potential Ref.~\cite{bartok10, bartok13}.
\label{fig:structures}}
\end{center}
\end{figure}

  As a first application of the package, we describe the automated
procedure for generating and relaxing a comprehensive array of grain boundary 
microscopic configurations. 
The generation of these job arrays is handled by \texttt{run\_gb\_net.py}. 
This script generates a set of inital microscopic structures for a given canonical grain boundary geometry, 
as required to approximately span all microscopic initializations at the boundary, 
and then calls the desired atomistic calculator in \texttt{quippy} to handle the structural minimisations. 
If the desired interatomic potential is not supported in the \texttt{quippy} package, 
the routines in \texttt{Imeall} will still generate the structures, 
which can then be manually forwarded to the calculator of choice. 
The relaxation method is customisable with the default set to be the FIRE algorithm of Ref.~\cite{bitzek06}. 
The structural relaxation will proceed until a desired force tolerance is achieved. 
A strain filter is attached to the atoms object during the structural minimisation so that the lattice 
vector orthogonal to the grain boundary plane can vary during the relaxation. 
This allows the structural minimisation to be carried out at zero pressure, or under any desired uniaxial load. 
  Fig.~\ref{fig:gb_energies} reports the grain boundary energies for the minimum 
energy structures obtained for the [001] orientation axis and a broad range of misorientation angles, 
using four different interatomic potentials. 
The interfacial energies are determined by a combination of the elastic strain energy and the 
chemical reconstructions that take place Ref.~\cite{read50}. 
In the low angle regime there is a clear distinction between the two contributions, 
corresponding to a picture of sections of coherent crystal populated by an array of
sparse, parallel dislocation cores.
As the dislocation core spacing approaches the lattice plane spacing, the decomposition 
into separated elastic and chemical contributions to the energy becomes more problematic. 
Fig.~\ref{fig:structures} illustrates the relaxed dislocation core structures associated 
with the same grain boundary using four different potentials. 
The local atomic environment of the atoms at the dislocation core are differentiated using the method of Ref.~\cite{ackland06}.
The significant structural variation obtained from the different potentials significantly affects the predicted
properties of the grain boundary. 
This motivates the construction of the present single repository. as a prerequisite to rationalize the differences.

The equilibrium spacing between dislocations also has a significant 
effect on the elastic properties of the interface. 
In low angle grain boundaries the dislocation spacings are governed by Frank's formula: 
\begin{equation}
d=\frac{|\mathbf{b}|}{2\sin(\Delta \theta/2)},
\end{equation}
where $\mathbf{b}$ is the Burgers vector and the angle $\theta$ is measured 
from the nearest coincident site lattice. 
Frank's formula readily allows to check whether or not the atomic potential is 
correctly describing the long range elastic strain field and the 
Burgers vector of isolated dislocations. 
\begin{figure}[htb]
\begin{center}
\includegraphics[width=\columnwidth]{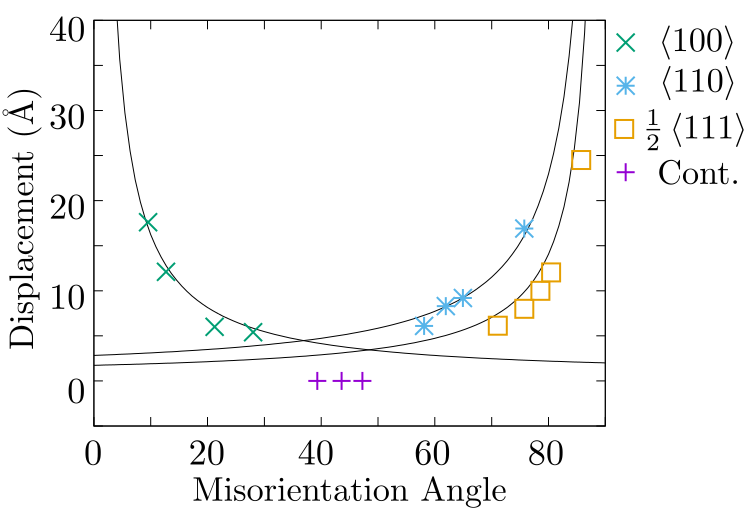}
\caption{Dislocation spacing in selected $[001]$ grain boundaries, with the unique 
Burger's vector as determined by the DXA (see text). 
The solid lines are determined by Frank's formula, the marked points correspond to the dislocation spacing determined
by atomistic relaxation, and the points marked 'Cont.' refer to large angle grain boundaries
where the dislocation cores merge into a continuous network. \label{fig:frankdiagram}}
\end{center}
\end{figure}
The computed dislocation spacings and Burgers vectors are compared with Frank's formula in Fig.~\ref{fig:frankdiagram}.
The dislocation character and spacing are determined using the DXA technique developed 
in Refs.~\cite{stukowskidxa10, stukowskidxa12}. 
The \texttt{Imeall} package also allows for the calculation of the Nye tensor using the 
technique described in Ref.~\cite{hartley05} to identify isolated dislocations. 
Knowledge of the structure of the dislocations and their spacing allows the 
construction of analytical models describing the elastic properties of interfaces. 
The accessibility of such data for a range of grain boundaries and force 
models provided further motivation for developing the \texttt{Imeall} database and tools.
\section{Interstitial Sites and Segregation Energies}
\label{sec:interstitials}
The possibility of cataloguing trapping sites and trap depths for interstitials is
a prerequisite for calculating the diffusivity and equilibrium concentrations
of point defects in a material: of particular interest for iron-based materials is hydrogen diffusivity\cite{hirth80}, 
notably in relation to the steel-embrittlement problem.  
A number of diffusivity models require parameterizations reliant on knowing trapping and 
segregation energies for boundaries \cite{kircheim82, kircheim88, du12, yamaguchi12}. 
Frequently the distribution of trap site energies is taken to be 
Gaussian with the variance and center value of the Gaussian fit to
experimental data \cite{kircheim88} but this so far had to be assumed 
rather than calculated with appropriate configuration space statistics.

\begin{figure}[htb]
\begin{center}
\includegraphics[width=0.8\columnwidth]{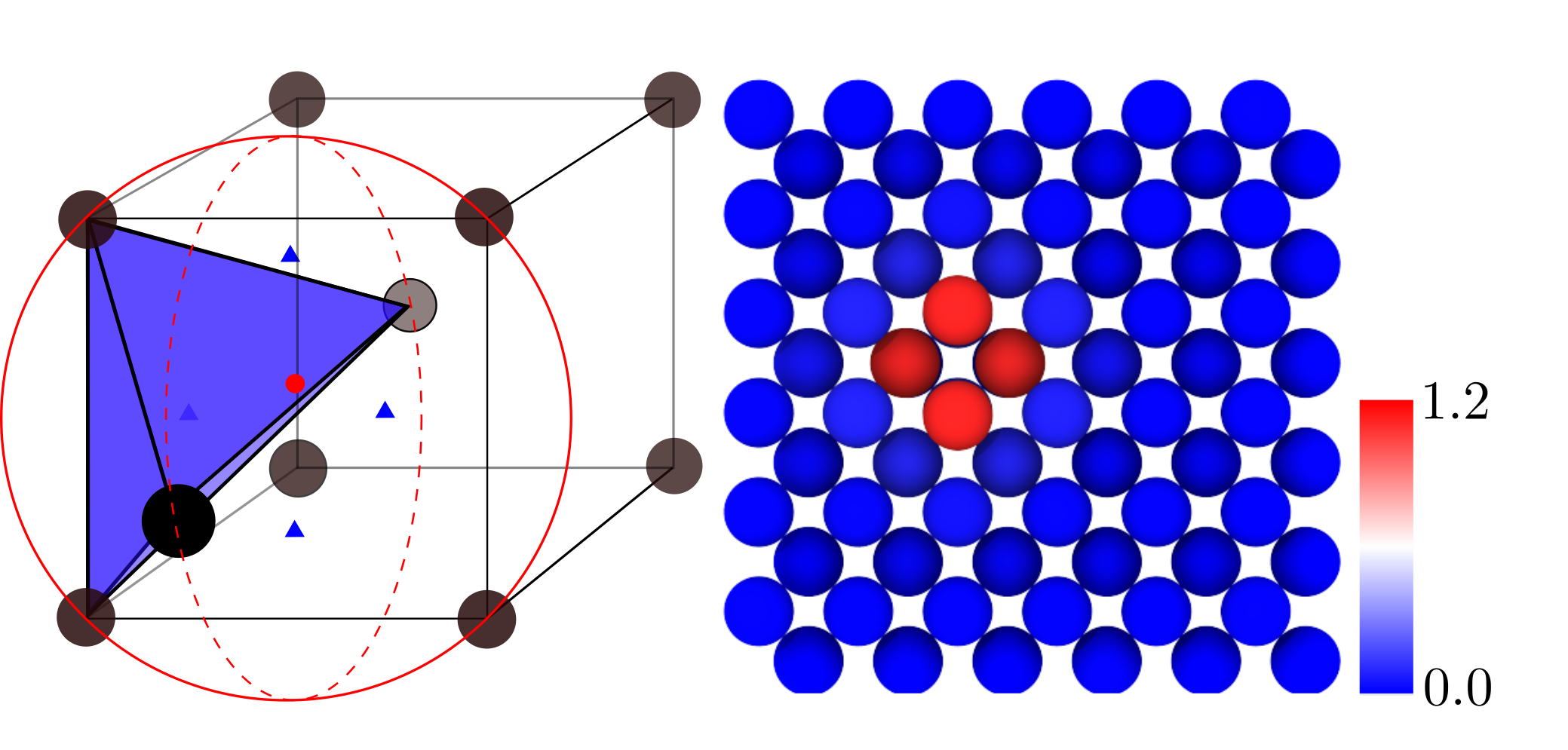}
\caption{
Automatic determination of interstitial sites in a bulk BCC lattice using the Delaunay triangulation method. 
In a bulk BCC lattice the octahedral sites can be determined as the 
circumcenter of the sphere represented with the solid and dashed red lines. 
The solid red circle is centred in an octahedral interstitial site (solid red dot), 
and the blue triangular faces comprise a tetrahedral site, 
located at the centre of an irregular tetrahedron having vertices on four lattice sites. 
The right panel provides the magnitude of the forces induced be a hydrogen point defect in eV/\AA units. 
These forces are required for calculation of the elastic dipole tensor.\label{fig:delaunaybulk}}
\end{center} 
\end{figure}

The \texttt{Imeall} database is equipped with the capability of cataloguing 
trapping sites and calculating the point defect interactions for all 
individual boundaries across the entire misorientation range and for different axes.
Indexing the possible segregation sites at the boundary and calculating realistic 
distributions for their associated trapping potentials is important for 
determining equilibrium interstitial occupancies at the interfaces\cite{veiga13}.
The interstitial sites in a bulk BCC lattice are represented schematically in Fig.~\ref{fig:delaunaybulk}.  
These sites can be determined automatically for any given 
atomistic structure using a Delaunay triangulation. 
Due to the intrinsic lattice distorsions induced by the geometric boundary 
between materials grains and its associated peculiar pattern of possible 
atomic relaxations, each grain boundary typically hosts a variety of 
non-equivalent interstital sites, differing in coordination and 
volume from the reference bulk lattice values.  
The routine \texttt{hydrogenate.py} contains the \texttt{Hydrogenate} class for indexing the interstitial sites 
in a lattice, decorating a lattice with hydrogen, and computing the 
volume of interstitial sites.
The routine \texttt{hydrogenate.py} contains the \texttt{Hydrogenate} class for indexing the interstitial sites 
in a lattice, decorating a lattice with interstitial hydrogens, and computing the volume of the interstitial sites. 
An alternative, and more complete, framework described in Ref.~\cite{goyal2017} which possesses extended functionality 
can also be used.

For each interstitial site of a given grain boundary it is possible to define an elastic dipole tensor (cf. Refs.~\cite{gillan84, freedman09, nazarov16}),  a local quantity of particular significance because it allows modelling the coupling  of a point 
defect to the strain fields of isolated dislocations, e.g., in the manner prescribed by Ref.~\cite{cochardt55}.
The elastic dipole tensor elements $G_{ij}$ are defined (cf. Ref.~\cite{freedman09}) as:
\begin{equation}
\label{eq:elasticdipole}
\frac{\partial \sigma_{ij}}{\partial n_{d}}\Bigr|_{\epsilon} = G_{ij},
\end{equation}
where $\sigma_{ij}$ is the volume averaged stress tensor, and $n_{d}$ is the point defect (e.g., hydrogen) concentration.
To compute these components we implemented in the \texttt{imeall.calc\_elast\_dipole} module the  ``defect force" scheme described in Ref.~\cite{nazarov16}. 
In this procedure, a point defect is introduced inside a unit cell of the 
host material, and the structure is relaxed to its equilibrium zero-forces geometry.  
The defect is at this point removed from the model system keeping 
all other atoms fixed, and the resulting forces on the ions 
previously surrounding the interstitial are calculated without allowing any relaxation. 
These forces and the ion position vectors can then be related to the dipole tensor via the formula:
\begin{equation}
G_{ij} = \sum_{m \neq d} f_{i}^{[m,d]}(d_{j}-m_{j}),
\end{equation}
where $\textbf{d},\textbf{m}$ are the position vectors of the defect and host atoms, 
respectively, $\textbf{f}^{[m,d]}$ is the force vector induced on the $m$-indexed atom by the 
removal of the $d$ defect, and the subscripts $i,j$ refer to spatial dimension components.
We note that the defect force method does not require a single-Hamiltonian description 
(or an explicit total energy expression) for  the system, so that forces 
can be computed using mixed (``embedding") schemes i.e., combinations of 
descriptive potentials in the region of near crystallinity and more 
accurate quantum mechanical models at or very close to the point defect, 
for instance using the `Learn On the Fly' scheme described in Ref.~\cite{csanyi04}.
The absolute magnitude of the defect forces induced by a hydrogen interstitial in the 
BCC lattice is illustrated in Fig.~\ref{fig:delaunaybulk}, right panel, 
where as expected the majority of the calculated effect is limited to the 
metal ions neighbouring the interstitial site analysed. 

The availability of the grain boundary database enables screening calculations 
on many defect geometries (e.g., distorted tetrahedral trapping sites 
located in the neighbourhood of the grain boundary) using fast classical potentials. 
It is therefore interesting to determine what accuracy level could be 
expected by the fast force models used in this screening step. 
We thus computed the dipole tensor for a tetrahedral site in bulk 
BCC Fe using an EAM force field \cite{ramasubramaniam09} and a reference DFT calculation. 
The computed dipole tensor for the EAM is:
\begin{equation}
\left(
\begin{array}{ccc}
 4.08 & 0 & 0 \\
 0 & 4.08 & 0 \\
 0 & 0 & 3.12  \\
\end{array}
\right){\rm eV},
\end{equation}
and for the density functional calculation, 
\begin{equation}
\left(
\begin{array}{ccc}
 4.65 & 0 & 0 \\
 0 & 4.65 & 0 \\
 0 & 0 & 3.52  \\
\end{array}
\right){\rm eV}.
\end{equation}
The DFT calculation was performed using the Vasp package, using the 
PBE functional approximation to treat exchange and correlation \cite{perdew77},
a PAW pseudopotential \cite{blochl94}, a $3\times3\times3$~k-point mesh and imposing a 45~Ry
energy cutoff on the plane wave wavefunction expansion.  
The periodic unit cell contained 250 Fe atoms with the H defect atom placed in a tetrahedral 
interstitial site in the middle of the unit cell.
These results suggest a ($\sim$ 12\%) underestimation of the tensor elements size 
by the EAM potential. 
Such first order error could have a significant impact on a theoretical analysis 
of the interaction of the point defect with an elastic strain field present in the metal matrix. 
However, such concern would be lifted by computing more precise values with higher accuracy DFT-based calculations 
only for a subset of most relevant/interesting cases revealed by an initial EAM-based high-throughput screening analysis, 
since systematic absolute $\sim$ 10\% errors could be tolerated, counting on error cancellations, 
in the initial screening used to identify the subset. 
The ability to catalogue quantities such as the elastic dipole tensor according to the 
potential used, and to identify artefacts resulting from the 
model (e.g., by comparing different models and pointing out outliers) is in fact 
a useful additional function of the \texttt{Imeall} package.
%
%
Similar calculations on the energetics of interstitials can be performed on 
any desired grain boundary structures and will be reported elsewhere.

\section{Conclusion}
We have described the structure and function of the \texttt{Imeall} package. 
The introduction of a naming convention and the overall structure of the database
and specification of the data models provides a very convenient 
framework for a computational resource relating to interfacial
structures in materials. 
The capabilities of the resource have been demonstrated with
reference to various properties of symmetric tilt boundaries and pure
crystalline $\alpha$-Fe. 
The resource is offered online as fully open-access and is extensible by any user. 
The code repository can be found at \url{https://github.com/kcl-tscm/imeall}, links to the full structure
database, which is hosted on the NOMAD servers and the web framework can be found in the
documentation of the package at \url{http://kcl-tscm.github.io/imeall/index.html}.

\section{Acknowledgements}
We would like to thank Prof. Adrian Sutton for illuminating conversations on 
grain boundaries and the physics of strain fields in their vicinity.  
We would also like to thank Dr. Thomas Daff and Prof. G. Cs\'anyi for making 
their Fe GAP potential available for comparison. 
The GAP software is available for non-commercial use at \texttt{www.libatoms.org}.
Financial support was provided by the Engineering and Physical Sciences Research Council under the HEmS program grant
EP/L014742/1 and grant EP/P002188/1. 
This research used resources of the Argonne Leadership Computing Facility,
which is a DOE Office of Science User Facility supported under 
Contract DE-AC02-06CH11357.
We are grateful to the UK Materials and Molecular Modelling Hub for computational resources, which is partially funded by EPSRC
(EP/P020194/1).
JRK, AF, and ADV acknowledge funding from the European Union’s Horizon 2020 research and innovation 
program (Grant No. 676580, The NOMAD Laboratory, a European Centre of Excellence).
\bibliographystyle{elsarticle-num}
\bibliography{gbrefs}

\begin{thebibliography}{10}
\expandafter\ifx\csname url\endcsname\relax
  \def\url#1{\texttt{#1}}\fi
\expandafter\ifx\csname urlprefix\endcsname\relax\def\urlprefix{URL }\fi
\expandafter\ifx\csname href\endcsname\relax
  \def\href#1#2{#2} \def\path#1{#1}\fi

\bibitem{vattre16}
A.~Vattr{\'e}, T.~Jourdan, H.~Ding, M.-C. Marinica, M.~J. Demkowicz,
  \href{http://dx.doi.org/10.1038/ncomms10424}{Non-random walk diffusion
  enhances the sink strength of semicoherent interfaces}, Nature Communications
  7 (2016) 10424, article.
\newline\urlprefix\url{http://dx.doi.org/10.1038/ncomms10424}

\bibitem{zhou17}
X.~Zhou, J.~Song,
  \href{http://www.sciencedirect.com/science/article/pii/S0167577X17303695}{Effect
  of local stress on hydrogen segregation at grain boundaries in metals},
  Materials Letters 196 (2017) 123 -- 127.
\newblock \href
  {http://dx.doi.org/http://dx.doi.org/10.1016/j.matlet.2017.03.035}
  {\path{doi:http://dx.doi.org/10.1016/j.matlet.2017.03.035}}.
\newline\urlprefix\url{http://www.sciencedirect.com/science/article/pii/S0167577X17303695}

\bibitem{bachurin10}
D.~Bachurin, D.~Weygand, P.~Gumbsch,
  \href{http://www.sciencedirect.com/science/article/pii/S1359645410003228}{Dislocation–grain
  boundary interaction in 〈1 1 1〉 textured thin metal films}, Acta
  Materialia 58~(16) (2010) 5232 -- 5241.
\newblock \href
  {http://dx.doi.org/http://dx.doi.org/10.1016/j.actamat.2010.05.037}
  {\path{doi:http://dx.doi.org/10.1016/j.actamat.2010.05.037}}.
\newline\urlprefix\url{http://www.sciencedirect.com/science/article/pii/S1359645410003228}

\bibitem{spearot14}
D.~E. Spearot, M.~D. Sangid,
  \href{http://www.sciencedirect.com/science/article/pii/S1359028614000175}{Insights
  on slip transmission at grain boundaries from atomistic simulations}, Current
  Opinion in Solid State and Materials Science 18~(4) (2014) 188 -- 195, slip
  Localization and Transfer in Deformation and Fatigue of Polycrystals.
\newblock \href
  {http://dx.doi.org/http://dx.doi.org/10.1016/j.cossms.2014.04.001}
  {\path{doi:http://dx.doi.org/10.1016/j.cossms.2014.04.001}}.
\newline\urlprefix\url{http://www.sciencedirect.com/science/article/pii/S1359028614000175}

\bibitem{bristowe75}
P.~D. Bristowe, A.~G. Crocker,
  \href{http://dx.doi.org/10.1080/14786437508226533}{A computer simulation
  study of the structures of twin boundaries in body-centred cubic crystals},
  Philosophical Magazine 31~(3) (1975) 503--517.
\newblock \href
  {http://arxiv.org/abs/http://dx.doi.org/10.1080/14786437508226533}
  {\path{arXiv:http://dx.doi.org/10.1080/14786437508226533}}, \href
  {http://dx.doi.org/10.1080/14786437508226533}
  {\path{doi:10.1080/14786437508226533}}.
\newline\urlprefix\url{http://dx.doi.org/10.1080/14786437508226533}

\bibitem{vitek80}
V.~Vitek, D.~A. Smith, R.~C. Pond,
  \href{http://dx.doi.org/10.1080/01418618008239340}{Structure of tilt grain
  boundaries in b.c.c. metals}, Philosophical Magazine A 41~(5) (1980)
  649--663.
\newblock \href
  {http://arxiv.org/abs/http://dx.doi.org/10.1080/01418618008239340}
  {\path{arXiv:http://dx.doi.org/10.1080/01418618008239340}}, \href
  {http://dx.doi.org/10.1080/01418618008239340}
  {\path{doi:10.1080/01418618008239340}}.
\newline\urlprefix\url{http://dx.doi.org/10.1080/01418618008239340}

\bibitem{tschopp07}
M.~Tschopp, G.~Tucker, D.~McDowell,
  \href{http://www.sciencedirect.com/science/article/pii/S1359645407001966}{Structure
  and free volume of 〈1 1 0〉 symmetric tilt grain boundaries with the e
  structural unit}, Acta Materialia 55~(11) (2007) 3959 -- 3969.
\newblock \href
  {http://dx.doi.org/http://dx.doi.org/10.1016/j.actamat.2007.03.012}
  {\path{doi:http://dx.doi.org/10.1016/j.actamat.2007.03.012}}.
\newline\urlprefix\url{http://www.sciencedirect.com/science/article/pii/S1359645407001966}

\bibitem{sutton95}
A.~P. Sutton, R.~W. Balluffi, Interfaces in Crystalline Materials, Oxford
  University Press, Great Clarendon Street, Oxford OX2 6DP, 1995.

\bibitem{handscomb58}
D.~C. Handscomb, On the random disorientation of two cubes, Canada J. Math. 10
  (1958) 85.

\bibitem{zeiner05}
P.~Zeiner, Symmetries of coincidence site lattices of cubic lattices,
  Zeitschrift f{\"u}r Kristallographie - Crystalline Materials. 220 (2005)
  915–925.

\bibitem{goldman10}
R.~Goldman,
  \href{http://dx.doi.org/10.2200/S00292ED1V01Y201008CGR013}{Rethinking
  quaternions}, Synthesis Lectures on Computer Graphics and Animation 4~(1)
  (2010) 1--157.
\newblock \href
  {http://arxiv.org/abs/http://dx.doi.org/10.2200/S00292ED1V01Y201008CGR013}
  {\path{arXiv:http://dx.doi.org/10.2200/S00292ED1V01Y201008CGR013}}, \href
  {http://dx.doi.org/10.2200/S00292ED1V01Y201008CGR013}
  {\path{doi:10.2200/S00292ED1V01Y201008CGR013}}.
\newline\urlprefix\url{http://dx.doi.org/10.2200/S00292ED1V01Y201008CGR013}

\bibitem{grimmer74}
H.~Grimmer, W.~Bollmann, D.~H. Warrington,
  \href{http://dx.doi.org/10.1107/S056773947400043X}{{Coincidence-site lattices
  and complete pattern-shift in cubic crystals}}, Acta Crystallographica
  Section A 30~(2) (1974) 197--207.
\newblock \href {http://dx.doi.org/10.1107/S056773947400043X}
  {\path{doi:10.1107/S056773947400043X}}.
\newline\urlprefix\url{http://dx.doi.org/10.1107/S056773947400043X}

\bibitem{grimmer84}
H.~Grimmer, \href{http://dx.doi.org/10.1107/S0108767384000246}{{The generating
  function for coincidence site lattices in the cubic system}}, Acta
  Crystallographica Section A 40~(2) (1984) 108--112.
\newblock \href {http://dx.doi.org/10.1107/S0108767384000246}
  {\path{doi:10.1107/S0108767384000246}}.
\newline\urlprefix\url{http://dx.doi.org/10.1107/S0108767384000246}

\bibitem{rittner96}
J.~D. Rittner, D.~N. Seidman,
  \href{http://link.aps.org/doi/10.1103/PhysRevB.54.6999}{〈110〉 symmetric
  tilt grain-boundary structures in fcc metals with low stacking-fault
  energies}, Phys. Rev. B 54 (1996) 6999--7015.
\newblock \href {http://dx.doi.org/10.1103/PhysRevB.54.6999}
  {\path{doi:10.1103/PhysRevB.54.6999}}.
\newline\urlprefix\url{http://link.aps.org/doi/10.1103/PhysRevB.54.6999}

\bibitem{spearot07}
D.~E. Spearot, M.~A. Tschopp, K.~I. Jacob, D.~L. McDowell,
  \href{http://www.sciencedirect.com/science/article/pii/S1359645406006537}{Tensile
  strength of 〈1 0 0〉 and 〈1 1 0〉 tilt bicrystal copper interfaces},
  Acta Materialia 55~(2) (2007) 705 -- 714.
\newblock \href
  {http://dx.doi.org/http://dx.doi.org/10.1016/j.actamat.2006.08.060}
  {\path{doi:http://dx.doi.org/10.1016/j.actamat.2006.08.060}}.
\newline\urlprefix\url{http://www.sciencedirect.com/science/article/pii/S1359645406006537}

\bibitem{tschopp07III}
M.~A. Tschopp, D.~L. Mcdowell,
  \href{http://dx.doi.org/10.1080/14786430701455321}{Asymmetric tilt grain
  boundary structure and energy in copper and aluminium}, Philosophical
  Magazine 87~(25) (2007) 3871--3892.
\newblock \href
  {http://arxiv.org/abs/http://dx.doi.org/10.1080/14786430701455321}
  {\path{arXiv:http://dx.doi.org/10.1080/14786430701455321}}, \href
  {http://dx.doi.org/10.1080/14786430701455321}
  {\path{doi:10.1080/14786430701455321}}.
\newline\urlprefix\url{http://dx.doi.org/10.1080/14786430701455321}

\bibitem{tschopp12}
M.~A. Tschopp, K.~N. Solanki, F.~Gao, X.~Sun, M.~A. Khaleel, M.~F. Horstemeyer,
  \href{http://link.aps.org/doi/10.1103/PhysRevB.85.064108}{Probing grain
  boundary sink strength at the nanoscale: Energetics and length scales of
  vacancy and interstitial absorption by grain boundaries in
  $\ensuremath{\alpha}$-fe}, Phys. Rev. B 85 (2012) 064108.
\newblock \href {http://dx.doi.org/10.1103/PhysRevB.85.064108}
  {\path{doi:10.1103/PhysRevB.85.064108}}.
\newline\urlprefix\url{http://link.aps.org/doi/10.1103/PhysRevB.85.064108}

\bibitem{campbell93}
G.~H. Campbell, S.~M. Foiles, P.~Gumbsch, M.~R\"uhle, W.~E. King,
  \href{https://link.aps.org/doi/10.1103/PhysRevLett.70.449}{Atomic structure
  of the (310) twin in niobium: Experimental determination and comparison with
  theoretical predictions}, Phys. Rev. Lett. 70 (1993) 449--452.
\newblock \href {http://dx.doi.org/10.1103/PhysRevLett.70.449}
  {\path{doi:10.1103/PhysRevLett.70.449}}.
\newline\urlprefix\url{https://link.aps.org/doi/10.1103/PhysRevLett.70.449}

\bibitem{paxton96}
A.~T. Paxton, \href{http://stacks.iop.org/0022-3727/29/i=7/a=003}{Atomic
  structure of metallic interfaces}, Journal of Physics D: Applied Physics
  29~(7) (1996) 1689.
\newline\urlprefix\url{http://stacks.iop.org/0022-3727/29/i=7/a=003}

\bibitem{ramasubramaniam09}
A.~Ramasubramaniam, M.~Itakura, E.~A. Carter,
  \href{http://link.aps.org/doi/10.1103/PhysRevB.79.174101}{Interatomic
  potentials for hydrogen in $\ensuremath{\alpha}$\char21{}iron based on
  density functional theory}, Phys. Rev. B 79 (2009) 174101.
\newblock \href {http://dx.doi.org/10.1103/PhysRevB.79.174101}
  {\path{doi:10.1103/PhysRevB.79.174101}}.
\newline\urlprefix\url{http://link.aps.org/doi/10.1103/PhysRevB.79.174101}

\bibitem{bitzek06}
E.~Bitzek, P.~Koskinen, F.~G\"ahler, M.~Moseler, P.~Gumbsch,
  \href{http://link.aps.org/doi/10.1103/PhysRevLett.97.170201}{Structural
  relaxation made simple}, Phys. Rev. Lett. 97 (2006) 170201.
\newblock \href {http://dx.doi.org/10.1103/PhysRevLett.97.170201}
  {\path{doi:10.1103/PhysRevLett.97.170201}}.
\newline\urlprefix\url{http://link.aps.org/doi/10.1103/PhysRevLett.97.170201}

\bibitem{mendelev03}
M.~I. Mendelev, S.~Han, D.~J. Srolovitz, G.~J. Ackland, D.~Y. Sun, M.~Asta,
  \href{http://dx.doi.org/10.1080/14786430310001613264}{Development of new
  interatomic potentials appropriate for crystalline and liquid iron},
  Philosophical Magazine 83~(35) (2003) 3977--3994.
\newblock \href
  {http://arxiv.org/abs/http://dx.doi.org/10.1080/14786430310001613264}
  {\path{arXiv:http://dx.doi.org/10.1080/14786430310001613264}}, \href
  {http://dx.doi.org/10.1080/14786430310001613264}
  {\path{doi:10.1080/14786430310001613264}}.
\newline\urlprefix\url{http://dx.doi.org/10.1080/14786430310001613264}

\bibitem{dudarev05}
S.~L. Dudarev, P.~M. Derlet,
  \href{http://stacks.iop.org/0953-8984/17/i=44/a=003}{A 'magnetic' interatomic
  potential for molecular dynamics simulations}, Journal of Physics: Condensed
  Matter 17~(44) (2005) 7097.
\newline\urlprefix\url{http://stacks.iop.org/0953-8984/17/i=44/a=003}

\bibitem{chamati06}
H.~Chamati, N.~Papanicolaou, Y.~Mishin, D.~Papaconstantopoulos,
  \href{http://www.sciencedirect.com/science/article/pii/S0039602806001701}{Embedded-atom
  potential for fe and its application to self-diffusion on fe(1 0 0)}, Surface
  Science 600~(9) (2006) 1793 -- 1803.
\newblock \href
  {http://dx.doi.org/http://dx.doi.org/10.1016/j.susc.2006.02.010}
  {\path{doi:http://dx.doi.org/10.1016/j.susc.2006.02.010}}.
\newline\urlprefix\url{http://www.sciencedirect.com/science/article/pii/S0039602806001701}

\bibitem{bartok10}
A.~P. Bart\'ok, M.~C. Payne, R.~Kondor, G.~Cs\'anyi,
  \href{https://link.aps.org/doi/10.1103/PhysRevLett.104.136403}{Gaussian
  approximation potentials: The accuracy of quantum mechanics, without the
  electrons}, Phys. Rev. Lett. 104 (2010) 136403.
\newblock \href {http://dx.doi.org/10.1103/PhysRevLett.104.136403}
  {\path{doi:10.1103/PhysRevLett.104.136403}}.
\newline\urlprefix\url{https://link.aps.org/doi/10.1103/PhysRevLett.104.136403}

\bibitem{bartok13}
A.~P. Bart\'ok, R.~Kondor, G.~Cs\'anyi,
  \href{https://link.aps.org/doi/10.1103/PhysRevB.87.184115}{On representing
  chemical environments}, Phys. Rev. B 87 (2013) 184115.
\newblock \href {http://dx.doi.org/10.1103/PhysRevB.87.184115}
  {\path{doi:10.1103/PhysRevB.87.184115}}.
\newline\urlprefix\url{https://link.aps.org/doi/10.1103/PhysRevB.87.184115}

\bibitem{read50}
W.~T. Read, W.~Shockley,
  \href{https://link.aps.org/doi/10.1103/PhysRev.78.275}{Dislocation models of
  crystal grain boundaries}, Phys. Rev. 78 (1950) 275--289.
\newblock \href {http://dx.doi.org/10.1103/PhysRev.78.275}
  {\path{doi:10.1103/PhysRev.78.275}}.
\newline\urlprefix\url{https://link.aps.org/doi/10.1103/PhysRev.78.275}

\bibitem{ackland06}
G.~J. Ackland, A.~P. Jones,
  \href{https://link.aps.org/doi/10.1103/PhysRevB.73.054104}{Applications of
  local crystal structure measures in experiment and simulation}, Phys. Rev. B
  73 (2006) 054104.
\newblock \href {http://dx.doi.org/10.1103/PhysRevB.73.054104}
  {\path{doi:10.1103/PhysRevB.73.054104}}.
\newline\urlprefix\url{https://link.aps.org/doi/10.1103/PhysRevB.73.054104}

\bibitem{stukowskidxa10}
A.~Stukowski, K.~Albe,
  \href{http://stacks.iop.org/0965-0393/18/i=8/a=085001}{Extracting
  dislocations and non-dislocation crystal defects from atomistic simulation
  data}, Modelling and Simulation in Materials Science and Engineering 18~(8)
  (2010) 085001.
\newline\urlprefix\url{http://stacks.iop.org/0965-0393/18/i=8/a=085001}

\bibitem{stukowskidxa12}
A.~Stukowski, V.~V. Bulatov, A.~Arsenlis,
  \href{http://stacks.iop.org/0965-0393/20/i=8/a=085007}{Automated
  identification and indexing of dislocations in crystal interfaces}, Modelling
  and Simulation in Materials Science and Engineering 20~(8) (2012) 085007.
\newline\urlprefix\url{http://stacks.iop.org/0965-0393/20/i=8/a=085007}

\bibitem{hartley05}
C.~Hartley, Y.~Mishin,
  \href{http://www.sciencedirect.com/science/article/pii/S1359645404007062}{Characterization
  and visualization of the lattice misfit associated with dislocation cores},
  Acta Materialia 53~(5) (2005) 1313 -- 1321.
\newblock \href
  {http://dx.doi.org/http://dx.doi.org/10.1016/j.actamat.2004.11.027}
  {\path{doi:http://dx.doi.org/10.1016/j.actamat.2004.11.027}}.
\newline\urlprefix\url{http://www.sciencedirect.com/science/article/pii/S1359645404007062}

\bibitem{hirth80}
J.~P. Hirth, \href{http://dx.doi.org/10.1007/BF02654700}{Effects of hydrogen on
  the properties of iron and steel}, Metallurgical Transactions A 11~(6) (1980)
  861--890.
\newblock \href {http://dx.doi.org/10.1007/BF02654700}
  {\path{doi:10.1007/BF02654700}}.
\newline\urlprefix\url{http://dx.doi.org/10.1007/BF02654700}

\bibitem{kircheim82}
R.~Kirchheim,
  \href{http://www.sciencedirect.com/science/article/pii/0001616082900037}{Solubility,
  diffusivity and trapping of hydrogen in dilute alloys. deformed and amorphous
  metals—ii}, Acta Metallurgica 30~(6) (1982) 1069 -- 1078.
\newblock \href
  {http://dx.doi.org/http://dx.doi.org/10.1016/0001-6160(82)90003-7}
  {\path{doi:http://dx.doi.org/10.1016/0001-6160(82)90003-7}}.
\newline\urlprefix\url{http://www.sciencedirect.com/science/article/pii/0001616082900037}

\bibitem{kircheim88}
R.~Kirchheim,
  \href{http://www.sciencedirect.com/science/article/pii/0079642588900102}{Hydrogen
  solubility and diffusivity in defective and amorphous metals}, Progress in
  Materials Science 32~(4) (1988) 261 -- 325.
\newblock \href
  {http://dx.doi.org/http://dx.doi.org/10.1016/0079-6425(88)90010-2}
  {\path{doi:http://dx.doi.org/10.1016/0079-6425(88)90010-2}}.
\newline\urlprefix\url{http://www.sciencedirect.com/science/article/pii/0079642588900102}

\bibitem{du12}
Y.~A. Du, J.~Rogal, R.~Drautz,
  \href{http://link.aps.org/doi/10.1103/PhysRevB.86.174110}{Diffusion of
  hydrogen within idealized grains of bcc fe: A kinetic monte carlo study},
  Phys. Rev. B 86 (2012) 174110.
\newblock \href {http://dx.doi.org/10.1103/PhysRevB.86.174110}
  {\path{doi:10.1103/PhysRevB.86.174110}}.
\newline\urlprefix\url{http://link.aps.org/doi/10.1103/PhysRevB.86.174110}

\bibitem{yamaguchi12}
M.~Yamaguchi, J.~Kameda, K.-I. Ebihara, M.~Itakura, H.~Kaburaki,
  \href{http://dx.doi.org/10.1080/14786435.2011.645077}{Mobile effect of
  hydrogen on intergranular decohesion of iron: first-principles calculations},
  Philosophical Magazine 92~(11) (2012) 1349--1368.
\newblock \href
  {http://arxiv.org/abs/http://dx.doi.org/10.1080/14786435.2011.645077}
  {\path{arXiv:http://dx.doi.org/10.1080/14786435.2011.645077}}, \href
  {http://dx.doi.org/10.1080/14786435.2011.645077}
  {\path{doi:10.1080/14786435.2011.645077}}.
\newline\urlprefix\url{http://dx.doi.org/10.1080/14786435.2011.645077}

\bibitem{veiga13}
R.~G.~A. Veiga, M.~Perez, C.~S. Becquart, C.~Domain,
  \href{http://stacks.iop.org/0953-8984/25/i=2/a=025401}{Atomistic modeling of
  carbon cottrell atmospheres in bcc iron}, Journal of Physics: Condensed
  Matter 25~(2) (2013) 025401.
\newline\urlprefix\url{http://stacks.iop.org/0953-8984/25/i=2/a=025401}

\bibitem{goyal2017}
A.~Goyal, P.~Gorai, H.~Peng, S.~Lany, V.~Stevanović,
  \href{http://www.sciencedirect.com/science/article/pii/S0927025617300010}{A
  computational framework for automation of point defect calculations},
  Computational Materials Science 130 (2017) 1 -- 9.
\newblock \href
  {http://dx.doi.org/https://doi.org/10.1016/j.commatsci.2016.12.040}
  {\path{doi:https://doi.org/10.1016/j.commatsci.2016.12.040}}.
\newline\urlprefix\url{http://www.sciencedirect.com/science/article/pii/S0927025617300010}

\bibitem{gillan84}
M.~J. Gillan, \href{http://stacks.iop.org/0022-3719/17/i=9/a=006}{The elastic
  dipole tensor for point defects in ionic crystals}, Journal of Physics C:
  Solid State Physics 17~(9) (1984) 1473.
\newline\urlprefix\url{http://stacks.iop.org/0022-3719/17/i=9/a=006}

\bibitem{freedman09}
D.~A. Freedman, D.~Roundy, T.~A. Arias,
  \href{http://link.aps.org/doi/10.1103/PhysRevB.80.064108}{Elastic effects of
  vacancies in strontium titanate: Short- and long-range strain fields, elastic
  dipole tensors, and chemical strain}, Phys. Rev. B 80 (2009) 064108.
\newblock \href {http://dx.doi.org/10.1103/PhysRevB.80.064108}
  {\path{doi:10.1103/PhysRevB.80.064108}}.
\newline\urlprefix\url{http://link.aps.org/doi/10.1103/PhysRevB.80.064108}

\bibitem{nazarov16}
R.~Nazarov, J.~S. Majevadia, M.~Patel, M.~R. Wenman, D.~S. Balint,
  J.~Neugebauer, A.~P. Sutton,
  \href{https://link.aps.org/doi/10.1103/PhysRevB.94.241112}{First-principles
  calculation of the elastic dipole tensor of a point defect: Application to
  hydrogen in $\ensuremath{\alpha}$-zirconium}, Phys. Rev. B 94 (2016) 241112.
\newblock \href {http://dx.doi.org/10.1103/PhysRevB.94.241112}
  {\path{doi:10.1103/PhysRevB.94.241112}}.
\newline\urlprefix\url{https://link.aps.org/doi/10.1103/PhysRevB.94.241112}

\bibitem{cochardt55}
A.~Cochardt, G.~Schoek, H.~Wiedersich,
  \href{http://www.sciencedirect.com/science/article/pii/0001616055901115}{Interaction
  between dislocations and interstitial atoms in body-centered cubic metals},
  Acta Metallurgica 3~(6) (1955) 533 -- 537.
\newblock \href
  {http://dx.doi.org/http://dx.doi.org/10.1016/0001-6160(55)90111-5}
  {\path{doi:http://dx.doi.org/10.1016/0001-6160(55)90111-5}}.
\newline\urlprefix\url{http://www.sciencedirect.com/science/article/pii/0001616055901115}

\bibitem{csanyi04}
G.~Cs\'anyi, T.~Albaret, M.~C. Payne, A.~De~Vita,
  \href{https://link.aps.org/doi/10.1103/PhysRevLett.93.175503}{``learn on the
  fly'': A hybrid classical and quantum-mechanical molecular dynamics
  simulation}, Phys. Rev. Lett. 93 (2004) 175503.
\newblock \href {http://dx.doi.org/10.1103/PhysRevLett.93.175503}
  {\path{doi:10.1103/PhysRevLett.93.175503}}.
\newline\urlprefix\url{https://link.aps.org/doi/10.1103/PhysRevLett.93.175503}

\bibitem{perdew77}
J.~P. Perdew, K.~Burke, M.~Ernzerhof,
  \href{https://link.aps.org/doi/10.1103/PhysRevLett.77.3865}{Generalized
  gradient approximation made simple}, Phys. Rev. Lett. 77 (1996) 3865--3868.
\newblock \href {http://dx.doi.org/10.1103/PhysRevLett.77.3865}
  {\path{doi:10.1103/PhysRevLett.77.3865}}.
\newline\urlprefix\url{https://link.aps.org/doi/10.1103/PhysRevLett.77.3865}

\bibitem{blochl94}
P.~E. Bl\"ochl,
  \href{https://link.aps.org/doi/10.1103/PhysRevB.50.17953}{Projector
  augmented-wave method}, Phys. Rev. B 50 (1994) 17953--17979.
\newblock \href {http://dx.doi.org/10.1103/PhysRevB.50.17953}
  {\path{doi:10.1103/PhysRevB.50.17953}}.
\newline\urlprefix\url{https://link.aps.org/doi/10.1103/PhysRevB.50.17953}

\end{thebibliography}


\begin{thebibliography}{0}
\bibitem{1}QUIPPY \url{https://libatoms.github.io/QUIP/}
\bibitem{2}ASE \url{https://wiki.fysik.dtu.dk/ase/} 
\bibitem{3}OVITO \url{https://www.ovito.org}       
\bibitem{4}Transformations \url{http://www.lfd.uci.edu/~gohlke/code/transformations.py}
\bibitem{5}pylada-defect \url{https://github.com/pylada/pylada-defects}
\end{thebibliography}
\end{document}